\documentclass[10pt,aps,prd,superscriptaddress,nofootinbib,nobibnotes,longbibliography,floatfix,preprintnumbers,twocolumn]{revtex4-2}

\usepackage{bm}
\usepackage{mathtools,
amsmath,
amssymb,
amsfonts,
mathrsfs,
chngcntr,
multirow}

\let\cc\corresponds
\let\corresponds\relax
\usepackage{mathabx}
\let\corresponds\cc

\usepackage[utf8]{inputenc}
\usepackage[T1]{fontenc}

\usepackage[dvipsnames]{xcolor}
\usepackage[unicode]{hyperref}
\hypersetup{colorlinks=true, citecolor=MidnightBlue,
            linkcolor=MidnightBlue, urlcolor=MidnightBlue, linktocpage=true}
\usepackage[normalem]{ulem}

\usepackage{graphicx}

\newcommand{\Z}{\mathbb{Z}}
\newcommand{\Mp}{M_{\rm Pl}}
\newcommand{\Msol}{M_\odot}

\renewcommand{\eqref}[1]{(\ref{#1})}
\newcommand{\figref}[1]{Fig.~\ref{#1}}

\begin{document}

\title{Constraining strongly-warped extra dimensions with  rotating black holes}

\author{Bruno Valeixo Bento}
\email{bruno.bento@ift.csic.es}
\affiliation{Instituto de Fisica Teórica, UAM-CSIC, C/ Nicolás Cabrera 13-15, Campus de Cantoblanco, 28049 Madrid Spain}
\preprint{IFT-25-151}

\author{Miquel Salicrú Herberg}
\email{miquel.salicru@estudiante.uam.es} 
\affiliation{Instituto de Fisica Teórica, UAM-CSIC, C/ Nicolás Cabrera 13-15, Campus de Cantoblanco, 28049 Madrid Spain}
\affiliation{Departamento de F\'isica Te\'orica, Universidad Aut\'onoma de Madrid, Cantoblanco, 28049 Madrid, Spain}

\begin{abstract}
    Massive bosonic fields can trigger superradiant instabilities in rotating astrophysical black holes leading to gaps in their mass-spin distribution. 
    For spin-2 fields, the instability timescale is orders of magnitude shorter than for any other superradiant mode, thereby yielding much stronger constraints. 
    We consider a tower of ultra-light spin-2 fields arising from a warped compactification of a single extra dimension and translate superradiant constraints on their masses into constraints on the warping. As a concrete scenario we consider the 2-brane Randall-Sundrum model and find constraints on the size of the extra dimension and the curvature of AdS$_5$. 
    We discuss the implications of these bounds for strongly warped throats and D-brane uplifts commonly used in attempts to realise metastable de Sitter vacua in string theory. 
\end{abstract}

\maketitle

\section{Introduction}

Whether fields beyond the Standard Model (SM) can be detected typically depends on two key properties: their mass and how strongly they couple to SM particles. This is the case for fifth-force constraints \cite{Adelberger:2003zx,Baker:2014zba,ParticleDataGroup:2024cfk}, coming from torsion table-top experiments \cite{Lee:2020zjt}, astrophysical tests \cite{Hardy:2016kme,Bottaro:2023gep,Fiorillo:2025zzx,Hardy:2025ajb}, atom interferometry \cite{Dimopoulos:2006nk}, the Event Horizon Telescope \cite{EventHorizonTelescope:2019dse,Psaltis:2018xkc} and collider searches \cite{Murata:2014nra}. Even though these experiments probe different regimes in different ways, they all depend on how strongly the new fields couple to the SM and become less sensitive for very weak coupling. 
Astrophysical rotating black holes (BHs) on the other hand might detect them \emph{regardless} of their weak coupling to SM particles, due to superradiant instabilities against massive bosonic perturbations \cite{Press:1972zz,Detweiler:1980uk,Cardoso:2004nk,Shlapentokh-Rothman:2013ysa}. Since superradiant instabilities are most efficient when the Compton wavelength of the particle is comparable to the gravitational radius of the BH, rotating BHs are amazing detectors of ultra-light fields---BHs with masses in the range $M\sim [1,10^{10}] \,M_\odot$ are sensitive to bosonic fields of mass $m_b\in [10^{-23},10^{-10}]~\text{eV}$ \cite{Arvanitaki:2009fg,Arvanitaki:2014wva,Brito:2015oca,Arvanitaki:2016qwi,Brito:2017wnc,Brito:2017zvb,Palomba:2019vxe,Isi:2018pzk,Brito:2020lup}. Massive tensor (spin-2) fields, in particular, were shown to trigger the strongest superradiant instabilities, with instability timescales much shorter than lower spin bosons \cite{Dias:2023ynv}.

Ultra-light fields arise both in bottom-up model building and top-down string theory (ST) constructions, where they are commonly linked to the extra dimensions of the UV theory---every massless field of the higher-dimensional theory gives rise to a number of massless fields and an infinite tower of massive modes upon compactification to a lower-dimensional Effective Field Theory (EFT). Among them there is always a massive spin-2 tower, arising from a massless graviton in higher dimensions. 
Detecting a massive Kaluza-Klein (KK) tower of spin-2 fields would therefore be a smoking gun for extra dimensions; these towers are constrained, for instance, by fifth force experiments, which rule out gravitationally coupled fields with masses $m_b \lesssim 5.1~\text{meV}$ \cite{Lee:2020zjt,ParticleDataGroup:2024cfk}. However, these fields may have couplings that differ from the gravitational one---either stronger or weaker than the Planckian coupling of the massless graviton---when the extra dimensions are warped, with the lower-dimensional scales and couplings depending on one's position along the warped directions. In particular, a warped compactification can lead to massive KK towers whose characteristic scale (i.e. mass gap) is exponentially suppressed compared to its unwarped counterpart, and whose couplings to SM fields can either be exponentially enhanced or exponentially suppressed depending on the exact realisation of the SM within the extra-dimensional setup. 

On the other hand, higher-dimensional rotating objects are known to suffer from instabilities \cite{Cardoso:2004zz,Cardoso:2005vk}. Rotating backgrounds of the form Kerr$_4\times \mathbb{R}^p$ are unstable against any massless mode for a wide range of wavelengths and frequencies in the transverse dimensions.\footnote{Interestingly, Kerr backgrounds in higher dimensions, Kerr$_d\times\mathbb{R}^p$ with $d>4$, are all stable against this mechanism, due to the lack of stable circular orbits in higher dimensional black hole spacetimes (or bound states of the effective potential) \cite{Cardoso:2004zz,Cardoso:2005vk}.} The instability can be avoided by compactifying the transverse dimensions at a scale smaller than the minimum wavelength for which the instability settles in---this is similar to the case of the Gregory-Laflamme (GL) instability \cite{Gregory:1993vy}, which is present for non-rotating black strings and branes as well. In fact, the instabilities can be interpreted as a superradiant instability of the 4d effective Kerr background in the presence of massive fields whose masses are given by the extra-dimensional momentum---in other words, they can be interpreted through dimensional reduction in terms of the massive KK towers in $4d$ \cite{Marolf:2004fya,Cardoso:2004zz,Cardoso:2005vk}. The requirement that the transverse dimensions are compactified to a small enough size is then translated into the requirement that all KK modes in the tower must be heavier than the minimum mass scale for which the superradiant instability occurs. 

In this work, we focus on the Randall-Sundrum two-brane model \cite{Randall:1999ee}, as a concrete example of a warped compactification, 
and show that constraints on the spectrum of massive spin-2 states \cite{Dias:2023ynv} from astrophysical BH observations can be used to set bounds on the parameters of the compactification and the scales that appear in the low-energy EFT. In light of these results, we discuss the case of warped throats and D-brane uplifts, commonly used in attempts to realise metastable de Sitter vacua in string theory \cite{Kachru:2003aw,Balasubramanian:2005zx,Conlon:2005ki}.

\section{Theoretical Background}
\subsection{Warped Compactifications}

Unwarped compactifications have higher-dimensional spacetimes that fully split into a 4d spacetime and a compact internal space, $M_D = M_4\times X_{D-4}$. Momentum along the compact directions is quantised and appears from the 4d point-of-view as the masses of states in an infinite tower. The intuition that very short distances should be hard to probe translates into the fact that these masses are typically much higher than the energy scales probed by our experiments---whenever this is the case, the massive tower can be consistently integrated out and its effects are encoded within a 4d EFT that contains irrelevant operators suppressed by the mass scale of the tower \cite{Burgess:2020tbq}. This scale is related to the volume of the compact internal space, 
\begin{equation}
    m_{\rm KK} \sim \mathcal{V}^{-\frac{1}{D-4}} \sim R^{-1}\,, 
\end{equation}
for a homogeneous space with characteristic size $R$, i.e. $\mathcal{V}\sim R^{D-4}$.

In a warped compactification, spacetime is a \emph{warped product} $M_D = M_4 \times_w X_{D-4}$ whose metric does not fully factorise,
\begin{equation}
    ds^2 = e^{2A(y)}g_{\mu\nu}dx^\mu dx^\nu + g_{mn}dy^m dy^n \,. 
    \label{eq:warped-metric}
\end{equation}
The exponential in front of the 4d metric is the \emph{warp factor} of the compactification, and its effect on 4d scales is $y$-dependent. This position dependence is important, for example, if the SM fields live on a lower-dimensional brane localised somewhere along the compact space, i.e. at fixed $y^m$.

\subsection{Randall-Sundrum Braneworlds}
\label{sec:RS}

\begin{figure}[t]
    \centering
    \includegraphics[width=\linewidth]{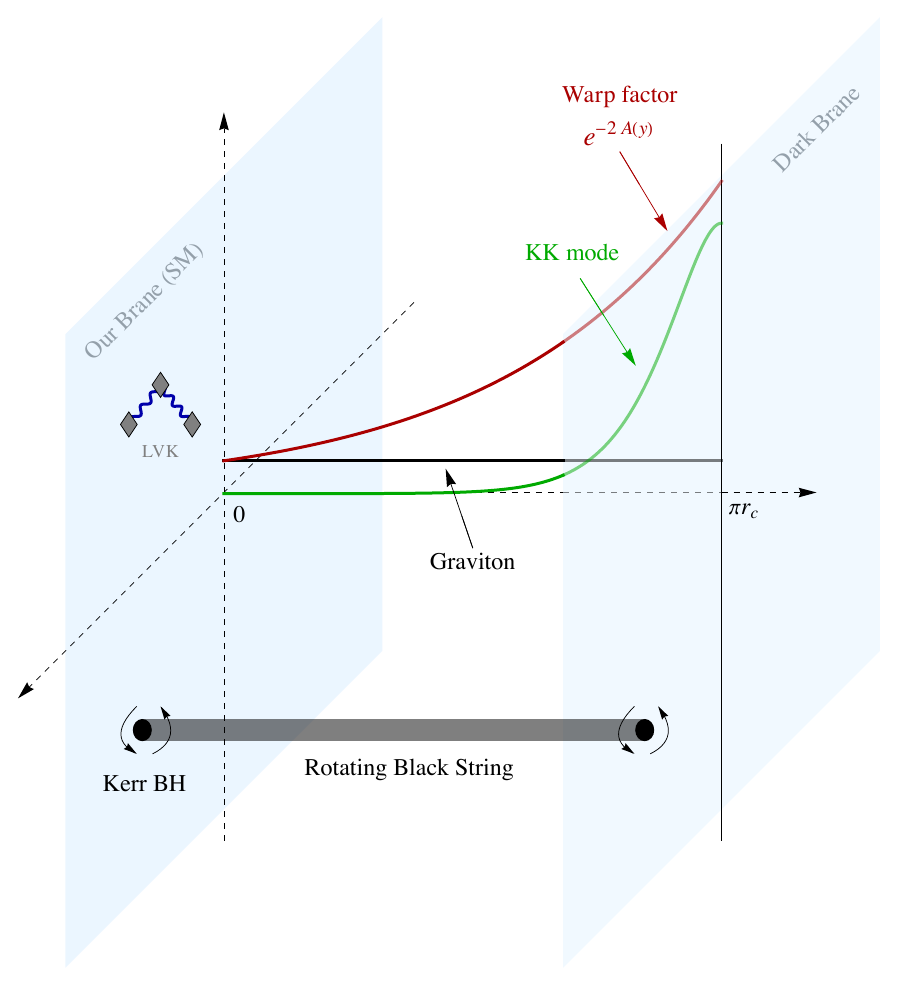}
    \caption{Pictorial representation of our 2-brane setup with a Standard Model brane and a dark brane at the boundaries of a warped interval of size $\pi r_c$. The warping localises the KK modes on the dark brane, while the graviton zero mode is homogeneous over the compact space. This localisation results in exponentially suppressed couplings (\figref{fig:RS-wavefunctions}) on the SM brane. We want to consider a 4d Kerr black hole, which could correspond to a 5d rotating black string stretching between the branes. 
    }
    \label{fig:RS-setup}
\end{figure}

The Randall-Sundrum model is a prototypical example of a warped extra-dimensional scenario \cite{Randall:1999ee,Randall:1999vf}. It consists of an AdS$_5$ background with one compact direction, an interval $S^1/\Z_2$ parametrised by the coordinate $0\leq y\leq\pi$. At each end point of the interval $(y=0,\pi)$ there is a 3-brane that can host a $(3+1)$-dimensional field theory (e.g. containing the SM). This setup is described by the action
\begin{align}
    S =& \int d^4x \int d y \,\sqrt{-G}\, \Big\{
        -\Lambda_5 + 2M_5^3 R_5 \nonumber \\
        &+ (\mathcal{L}_{1} - V_{1})\frac{\delta( y)}{\sqrt{G_{ y y}}} 
        + (\mathcal{L}_{2} - V_{2})\frac{\delta( y-\pi)}{\sqrt{G_{ y y}}}
    \Big\} \,, 
\end{align}
where $\Lambda_5 < 0$ is the 5d cosmological constant, $M_5$ the 5d Planck scale, and $\mathcal{L}_{i}/V_{i}$ are the matter Lagrangian/vacuum energy of each 3-brane \cite{Randall:1999ee,Randall:1999vf}. The equations of motion that follow from this action are solved by the metric
\begin{equation}
    ds^2 = e^{-2kr_c| y|}\eta_{\mu\nu} dx^\mu dx^\nu + r_c^2 d y^2 \,,
    \label{eq:RS-metric}
\end{equation}
where $\eta_{\mu\nu}$ is Ricci flat, $r_c$ is the radius of the $S^1$ prior to the $\Z_2$ orbifolding, independent of $ y$, so that the size of the interval is $\pi r_c$, and $k$ is a scale given by 
\begin{equation}
    k \equiv \sqrt{\frac{-\Lambda_5}{24M_5^3}} \,,
\end{equation}
in terms of which the brane vacuum energies are 
\begin{equation}
    V_{1} = -V_{2} = 24M_5^3 k \,.
\end{equation}
The metric in \eqref{eq:RS-metric} takes precisely the form of a warped compactification \eqref{eq:warped-metric},
with a 4d Planck scale 
\begin{equation}
    M_{\rm Pl}^2 = \frac{M_5^3}{k}(1 - e^{-2\pi\,kr_c}) \,,
    \label{eq:4d-Planck-scale}
\end{equation}
that depends weakly on the size of the interval whenever the warping is strong ($kr_c>1$). In this regime, the scales are essentially related by the characteristic AdS$_5$ curvature scale. 

On the other hand, a mass scale $m_0$ on the brane at $ y = \pi$ in the 5d theory corresponds to a physical mass
\begin{equation}
    m = e^{-\pi\,kr_c}m_0 \,,
    \label{eq:physical-mass}
\end{equation}
when measured with the 4d metric on the brane written in Einstein frame (i.e. corresponding to a canonical Einstein-Hilbert action in the 4d EFT) \cite{Randall:1999ee}. The key consequence of this setup is that a modest value of $kr_c\approx 12$ would be enough to connect the Planck scale with the TeV scale, thereby addressing the hierarchy problem without any large hierarchy between fundamental parameters \cite{Randall:1999ee}. 

Not only are the KK mode masses exponentially suppressed by warping, but so are the couplings to SM fields either suppressed or enhanced by the warp factor, depending on which brane hosts the SM. The eigenvalue problem that determines the KK spectrum in this setup is given by \cite{Randall:1999ee,Randall:1999vf,Boos:2012zz}
\begin{subequations}
    \begin{align}
    &(\partial_ y^2 - 4k^2r_c^2 - m^2r_c^2 e^{2k r_c y} 
    )\,\tilde{\xi} = 0 \\
    &\tilde{\xi}'(y) = -2kr_c\,\tilde{\xi}(y) \quad\text{at}\quad y=0,\pi  \,,
    \label{eq:KK-spectrum}
\end{align}
\end{subequations}
and it leads to the masses  
\begin{equation}
    m_n^{\rm (RS)} \approx \gamma_n\,k\, e^{-\pi\,kr_c} \,,
    \label{eq:RS-mass}
\end{equation}
with $\gamma_n$ approximately given by the roots of the Bessel function $J_1(\gamma_n)$ for small enough $n$. The wavefunctions of the 4d fluctuations $G_{\mu\nu}\to e^{-2kr_c |y|}(\eta_{\mu\nu}+h_{\mu\nu}\,\xi(y))$ are then
\begin{align}\label{eq:KK-wavefunctions}
    &\xi_n(y) = \mathcal{N}_n \, e^{2kr_c\,|y|} \\ 
    &\quad\quad\times\left[Y_1\left(\tfrac{m}{k}\right) J_2\left(e^{kr_c |y|}\tfrac{ m}{k}\right)-J_1\left(\tfrac{m}{k}\right)
   Y_2\left(e^{k r_c|y|}\tfrac{ m}{k}\right)\right] \,, \nonumber 
\end{align}
where $J_i\,,\,Y_i$ are Bessel functions and the normalisation factors $\mathcal{N}_n$ are such that 
\begin{equation}
    \int_{-\pi}^\pi dy\,e^{2kr_c\,|y|}\,\xi_i(y)\,\xi_j(y) = \delta_{ij} \,.
\end{equation}

\begin{figure}[t]
    \centering
    \includegraphics[width=0.95\linewidth]{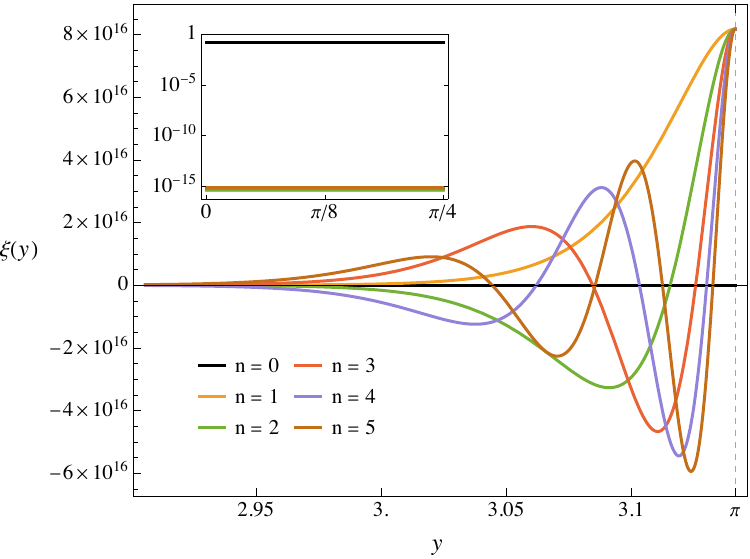}
    \caption{Kaluza-Klein mode wavefunctions for the Randall-Sundrum model \eqref{eq:KK-wavefunctions} normalised with respect to the graviton zero mode (for $kr_c=12$). The massive mode $(n>0)$ wavefunctions are localised and exponentially enhanced at $y=\pi$, and exponentially suppressed at $y=0$ .}
    \label{fig:RS-wavefunctions}
\end{figure}

The coupling strength of a KK graviton to a SM field $\Psi$ confined to the 3-brane at $y=y_{\rm SM}$ is proportional to the wavefunction of the KK mode evaluated at the position of the SM brane,
\begin{equation}
    h_{\mu\nu}^{(n)}(x,y)\Psi(x)\overline{\Psi}(x) \sim \xi_n(y_{\rm SM})\, h_{\mu\nu}^{(n)}\Psi\overline{\Psi} \,.
\end{equation} 
In the original RS proposal \cite{Randall:1999ee}, $y_{\rm SM}=\pi$ where warping is the strongest and the KK modes are localised (\figref{fig:RS-wavefunctions}), so that the couplings between KK modes and SM fields are exponentially \emph{enhanced},
\begin{equation}
    \xi_n(\pi) \sim e^{kr_c\,\pi} \sim 10^{16} \,,\quad\text{for}~kr_c\sim 12\,,
\end{equation}
which could make them detectable in current collider experiments. No evidence of KK modes has been found so far at the LHC, which itself puts constraints on the original RS proposal \cite{ParticleDataGroup:2024cfk}.

As we will see below, masses of order TeV do not fall within the superradiant instability window. Furthermore, we are interested in the case where the KK gravitons couple very weakly to SM particles, making their detection through fifth-force experiments extremely challenging. We therefore take the SM to be hosted on the brane at $y=0$, 
\begin{equation}
    \xi_n(0) \sim e^{-kr_c\,\pi} \,,
\end{equation}
so that couplings are exponentially suppressed, and consider larger values of $kr_c$ that bring the KK masses below the $10^{-11}$ eV instability threshold.

\subsection{Fifth-force constraints}

Constraints on fifth-forces are often combined through a common parameterisation in terms of a Yukawa-type correction to the Newtonian potential felt by two masses, 
\begin{equation}
    V(r) = \frac{Gm_1m_2}{r}\left\{1 + \alpha\,e^{-r/\lambda}\right\} \,,
\end{equation}
where $\alpha$ is a dimensionless parameter that encodes the coupling strength of the new field compared to the graviton coupling and $\lambda$ is the range of the interaction, related to the mass of the field as $\lambda = 1/m_b$. For example, collider experiments probe very short ranges ($m_b\lesssim$ TeV) but require couplings much stronger than gravitational ($\sqrt{\alpha}\gtrsim 10^{16}$); for comparison, torsion balance experiments can probe gravitational couplings ($\alpha\sim 1$) but are only sensitive up to $\lambda\gtrsim 50\,\mu m$ ($m_b\lesssim $ meV). These bounds take a prominent role in the context of the Dark Dimension (DD) scenario \cite{Montero:2022prj}, where a single large and unwarped extra dimension is motivated through Swampland considerations and experimental constraints. This model has lead to several predictions and attempts to connect string theory with current cosmological observations \cite{Gonzalo:2022jac,Anchordoqui:2022txe,Anchordoqui:2022svl,Anchordoqui:2023tln,Obied:2023clp,Anchordoqui:2024akj,Anchordoqui:2024dxu,Bedroya:2025fwh}. From our discussion here, it is clear that the absence of warping in the DD is crucial in its use of fifth-force constraints in determining the relevant scales; moreover, the scale of the DD makes the associated KK tower too heavy to be probed through superradiance.

\begin{figure}[!t]
    \centering
    \includegraphics[width=\linewidth]{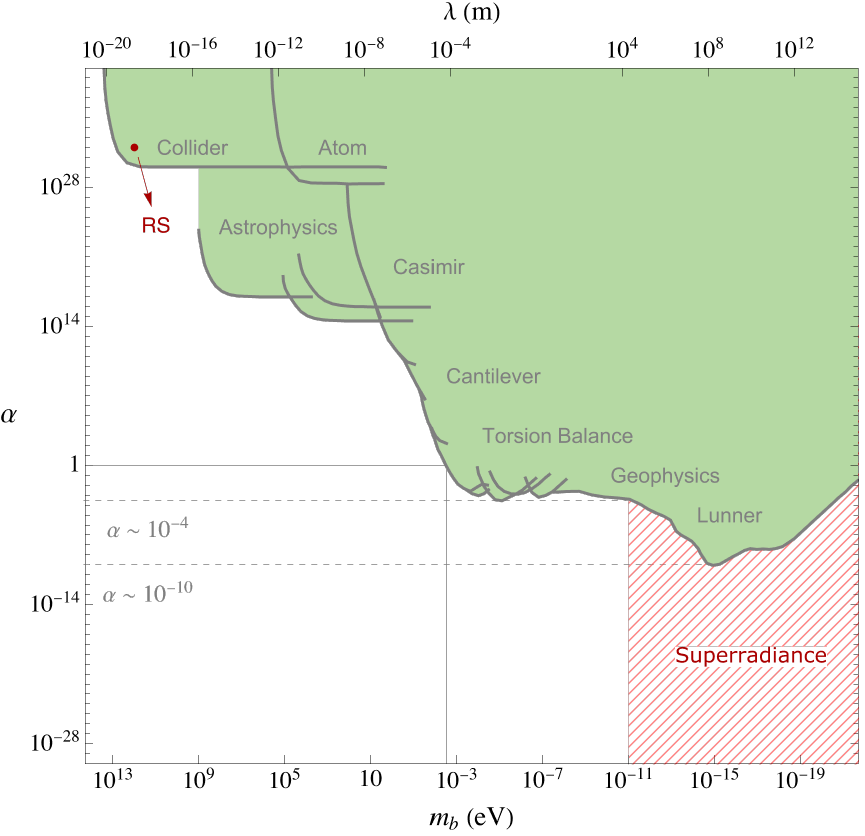}
    \caption{Experimental constraints on the parameters $\alpha$ (coupling strength) and $\lambda$ (range) of a Yukawa-type interaction, with the shaded area indicating the excluded region of parameter space at 95\% confidence level. Figure adapted from \cite{Murata:2014nra,Cembranos:2017vgi}. See \cite{Murata:2014nra,Cembranos:2017vgi} for details and further references.
    We also include the region of parameter space excluded by superradiant instabilities triggered by massive spin-2 modes with masses in the range $m_b\in(10^{-23},10^{-11})$ eV, from \cite{Dias:2023ynv}.}
    \label{fig:fifth-force-constraints}
\end{figure}

When the fifth-force is mediated by the lightest KK mode in our warped compactification, these parameters are given by 
\begin{align}
    \alpha\sim |\xi_1(0)|^2 \sim e^{-2kr_c\,\pi} 
    && \lambda \sim k^{-1} e^{kr_c\,\pi} \,,
\end{align}
and we can compare the excluded regions of parameter space due to fifth-force experiments with the fundamental parameters of the compactification $k,r_c$ (\figref{fig:fifth-force-constraints}). When this mode is heavy enough to be suppressed by the Yukawa exponential $e^{-r/\lambda}$, and since the coupling strength of different KK modes is similar, considering only the first mode is a reasonable approximation; on the other hand, if strong warping leads to several KK modes with $e^{-r/\lambda}\sim 1$, these add up as corrections only suppressed by the coupling $\alpha$. For this reason, for exponentially light KK modes, one should be careful in interpreting the bounds in \figref{fig:fifth-force-constraints}. In particular, a consistent interpretation of these bounds should take into account both the couplings and multiplicity of light KK modes (see e.g \cite{ValeixoBento:2022qca,Hardy:2025ajb})

For our purposes here, however, there is one key aspect of fifth-force constraints that is clear from \figref{fig:fifth-force-constraints}: these experiments are most sensitive to larger couplings. For fields with couplings to the SM that are weaker than Planckian, a large range of small masses is allowed by these experiments. In contrast, as we review in Section~\ref{sec:superradiance}, superradiant instabilities of rotating BHs are purely gravitational and do not depend on the couplings of new fields to the SM. In the context of \figref{fig:fifth-force-constraints}, superradiant instabilities and BH measurements have the potential to exclude a vertical band in the range $m_b\in(10^{-23},10^{-11})$~eV, regardless of the boson couplings to the SM. 

\subsection{Warped throats in string compactifications}

The RS setup we consider, with its warped extra dimension, can be seen as a lower-dimensional/low-energy proxy of a common feature of string constructions---the presence of strongly warped throats in the compact space.
Warped throats are common ingredients of string theory compactifications \cite{Klebanov:2000hb,Polchinski:1998rr,Ibanez:2012zz} and have been mostly studied in Type IIB superstring theory in the context of cosmological accelerated expansion \cite{Giddings:2001yu,Kachru:2003aw,Kachru:2003sx,Balasubramanian:2005zx,Conlon:2005ki,Cicoli:2023opf}. More precisely, the extra dimensions of string theory can be deformed by fluxes or localised sources, and develop highly-warped regions---warped throats---within which scales are exponentially suppressed with respect to the fundamental (bulk) scales. In models that attempt to construct de Sitter vacua \cite{Kachru:2003aw,Balasubramanian:2005zx,Conlon:2005ki} or produce inflation through brane-anti-brane potentials \cite{Kachru:2003sx,Cicoli:2024bwq}, an anti-D3 brane sits at the tip of the warped throat in such a way that its tension of order $M_s/g_s$ is suppressed and can compete with other terms in the scalar potential without overwhelming them. 

In these scenarios the SM brane(s) are expected to be localised away from the highly warped throat (typically corresponding to D7-brane stacks wrapping certain cycles in the bulk). Since these constructions do not try to address the electroweak hierarchy problem, one is not forced to fix the warp factor through the ratio $\Mp/M_{\rm ew}\sim 10^{15}$, which gives us the necessary freedom to address a different hierarchy, e.g. $M_s/V_{\rm AdS}$. This situation is thus analogous to the regime we consider in this paper---the tip of the conifold acts as the dark brane in \figref{fig:RS-setup}, while the SM will be hosted on some brane configuration localised away from the tip and towards the bulk, where these couplings are exponentially suppressed, allowing ultra-light KK modes to evade fifth force constraints \cite{ValeixoBento:2022qca}. 

The warped throat is usually described through the Klebanov-Strassler solution \cite{Klebanov:2000hb}, a warped deformed conifold sourced by 3-form fluxes with a warp factor given by \cite{Bento:2021nbb,ValeixoBento:2022qca}
\begin{equation}
    e^{2A_{\rm tip}} \approx c'\frac{e^{-\frac{2\pi K}{3g_sM}}}{(g_sM)\,\mathcal{V}^{1/3}} \,,
\end{equation}
at the tip of the throat, and $e^{2A_{\rm far}}\approx 1$ far from the tip, while the masses of the KK modes are given by \cite{Firouzjahi:2005qs,Shiu:2007tn,ValeixoBento:2022qca} 
\begin{equation}
    m_n 
    \sim \tilde{\gamma}_n\, \frac{e^{-\frac{4\pi K}{3g_sM}}}{g_sM}\,\frac{g_s}{\mathcal{V}^{1/3}}\, 
    M_p \,.
    \label{eq:KS-mass}
\end{equation}
Here $g_s$ is the string coupling, $K,\,M$ are the flux quanta sourcing the warping, $\mathcal{V}$ is the volume of the compact space in string units and $c'\approx 37$.
The effect of these warped spin-2 modes was compared with existing fifth force constraints in \cite{ValeixoBento:2022qca}, taking into account the coupling suppression caused by the wavefunction profiles away from the tip of the conifold. 

In the string theory case, the parameters are theoretically constrained as well: the flux quanta $(M,K)$ must be integers, $g_s\ll 1$ for control of the string loop perturbative expansion, and $g_sM > 1$ for control of the supergravity approximation \cite{Klebanov:2000hb}, which also requires $\mathcal{V}\gg 1$. Large values of $MK$ require objects with a large negative D3-brane charge in order to cancel tadpoles; this can be achieved in the perturbative regime, for example through Whittney branes \cite{Crino:2020qwk}. As an example, for 
$$g_s=0.33\,,\,M=6\,,\,K=28\,,\,\mathcal{V}=10^{10}\,,$$
we have $m_{\rm KK}\sim 10^{-11}$ eV, at the upper bound of the superradiance range, and a warp factor at the tip $e^{2A_{\rm tip}}\sim 10^{-15}$. 

Although several control issues related to warped compactifications and moduli stabilisation---in particular in connection with the anti-D3 brane uplifts common in de Sitter constructions---have been raised \cite{Bena:2018fqc,Bena:2020xrh,Gao:2020xqh,Plauschinn:2021hkp,Lust:2021xds,Gao:2022fdi,Gao:2022uop,Hebecker:2022zme,Grana:2022dfw,Lust:2022mhk,Lust:2022xoq,Becker:2024ayh,Schreyer:2022len,Schreyer:2024pml}, in Section \ref{sec:conclusions} we interpret the bounds found in Section \ref{sec:superradiance} in the context of the warped deformed conifold and show that they independently constrain the strength of the warping in this background.

\section{Superradiant Instabilities}
\label{sec:superradiance}

Rotating BHs have been shown to suffer from superradiant instabilities against massive bosonic perturbations \cite{Press:1972zz,Detweiler:1980uk,Cardoso:2004nk,Shlapentokh-Rothman:2013ysa}. These instabilities are triggered by the combination of superradiance effects---the amplification of bosonic perturbations within the ergoregion of a rotating BH---and the confining potential created by the mass of the bosonic field, which leads to an exponential growth of these perturbations. Mode amplification of the bosonic field occurs when the angular velocity of the BH horizon exceeds the angular phase velocity of the wave mode; thus there is an instability whenever the superradiance condition is satisfied,
\begin{equation}
    \omega_R < m\,\Omega_H \,,
    \label{eq:superradiance-condition}
\end{equation}
where $\omega_R$ is the frequency of the perturbation, $\Omega_H$ the horizon angular velocity and $m$ the azimuthal number of the unstable mode \cite{Brito:2015oca}.
In the superradiant regime the BH spins
down, transferring energy and angular momentum to a boson condensate until the superradiant condition is saturated, at which point true bound states may form \cite{Herdeiro:2014goa}.

The superradiant instability triggered by a boson with mass $m_b$ is most effective when its Compton wavelength is comparable to the BH size, i.e. for $\alpha\equiv m_bM\sim O(0.1)$
corresponding to $m_b\sim 10^{-11}(\Msol/M)$ eV \footnote{It is somewhat unfortunate that this coupling parameter is also conventionally called $\alpha$; this should not be confused with the parameter $\alpha$ discussed above. Moreover, this coupling is always discussed in units with $G=c=1$, while superradiant instabilities are typically discussed with $m_b$ in eV and $M$ in units of $\Msol$; restoring the appropriate units, the dimensionless coupling is $\alpha \approx 7.5\times 10^9\,\frac{m_b}{\rm eV}\frac{M}{\Msol}$.}. 
The instability can be triggered by bosons of various spins, including scalar, vector and tensor fields; while the scalar and vector cases have been much studied and understood \cite{Brito:2015oca}, the progress in fully understanding the spin-2 case is more recent \cite{Brito:2013wya,Brito:2013yxa,Brito:2020lup,Dias:2023ynv}. It turns out that the timescale of the dominant spin-2 unstable mode is much shorter than that of scalar and vector bosons, leading to much wider mass ranges capable of triggering an instability. 

Since the compactification of extra dimensions always leads to a KK tower of spin-2 states, this stronger spin-2 instability can lead to stronger bounds on the masses of these states. One can then interpret these bounds in terms of the fundamental parameters of the higher-dimensional theory, the compactification manifold and the ingredients used in the construction. Due to the ultra-light regime constrained by superradiance, only compactifications with ultra-light KK towers can be probed in this way; this then motivates us to explore the case of strongly-warped extra dimensions, with exponentially suppressed KK masses, as explained in the previous section. 

\subsection{Spin-2 dominant instability}

Massive tensor (spin-2) perturbations are less
understood than scalar and vector perturbations. 
In \cite{Dias:2023ynv} the superradiant instability
of a Kerr BH against massive spin-2 fields was finally computed without approximations and for any mode, completing the prior semi-analytical perturbative approach to first order in the spin \cite{Pani:2013pma,Brito:2013wya} and analytical results in the Newtonian regime of small gravitational couplings \cite{Brito:2020lup}. 
This was important because these approximations become inaccurate for most astrophysical BHs, on the one hand due to their considerable spins, and on the other since the most relevant regime for the instability is for $\mathcal{O}(0.1)$ couplings \cite{Dias:2023ynv}. Moreover, a ``special'' dipole mode was numerically found in \cite{Brito:2013wya} and conjectured to be the dominant one, with an instability timescale much shorter than all other unstable modes, but its full implications were only understood with the results of \cite{Dias:2023ynv}.

A spin-2 perturbation $H_{\mu\nu}$ of mass $m_b$ propagating on a Ricci-flat background is described by the field equations \cite{Brito:2013wya,Mazuet:2018ysa,Dias:2023ynv}
\begin{equation}
    \Box H_{\mu\nu} + 2R_{\mu\nu\rho\sigma}H^{\rho\sigma} - m_b^2 H_{\mu\nu} = 0 \,,
    \label{eq:spin-2-equations}
\end{equation}
and subject to the constraints
\begin{equation}
    \nabla^\mu H_{\mu\nu} = 0\,,
    \quad H^\mu_{\phantom{\mu}\mu} = 0 \,,
    \label{eq:spin-2-constraints}
\end{equation}
where $\Box\equiv\nabla^\mu\nabla_\mu$ is the d'Alembert operator in 4d and $R_{\mu\nu\rho\sigma}$ the Riemann tensor of the background. 

In this work we assume that there exists a background solution of the form 
\begin{equation}
    ds_5^2 = e^{2A(y)}ds^2_\text{4d Kerr} + r_c^2\,dy^2 \,,
\end{equation}
which corresponds to a warped rotating black string \cite{Seahra:2004fg,Bohra:2023vls}\footnote{Alternatively, one might want to consider a 5d black hole localised on the brane \cite{Shiromizu:1999wj,Sasaki:1999mi,Harko:2004ui,Aliev:2006qp,Estrada:2025ice, Kumar:2025jsi}. This might be particularly interesting in the case of the non-compact Randall-Sundrum II, for which fully backreacted rotating BH solutions were found in \cite{Biggs:2021iqw}.}.
The mass of the spin-2 field then follows from the Kaluza-Klein momentum along the warped direction $y$, as explained in Section~\ref{sec:RS} and explicitly given by \eqref{eq:RS-mass}. After dimensional reduction, we end up with a 4d Kerr background and a KK tower of massive spin-2 fields.
We refer to Appendix~\ref{ap:Kerr} for more details on the Kerr geometry and the setup of the eigenvalue problem that describes superradiance (see also \cite{Brito:2015oca,Dias:2023ynv}), and discuss here only the key steps.
The mass of the spin-2 field then follows from the Kaluza-Klein momentum along the warped direction $y$, as explained in Section~\ref{sec:RS} and explicitly given by \eqref{eq:RS-mass}. After dimensional reduction, we end up with a 4d Kerr background and a KK tower of massive spin-2 fields.
We refer to Appendix~\ref{ap:Kerr} for more details on the Kerr geometry and the setup of the eigenvalue problem that describes superradiance (see also \cite{Brito:2015oca,Dias:2023ynv}), and discuss here only the key steps.

In the usual Boyer–Lindquist coordinates $(t,r,\vartheta,\varphi)$, symmetries of the Kerr spacetime allow for the Fourier mode decomposition of the spin-2 field
\begin{equation}
    H_{\mu\nu} (t,r,\vartheta,\varphi) = e^{-i\omega t}e^{im\varphi}\tilde{H}_{\mu\nu}(r,\vartheta) \,,
\end{equation}
where $m$ is the azimuthal quantum number and $\omega$ the frequency of the perturbation, reducing the field equations to a coupled system of ten PDEs for $\tilde{H}_{\mu\nu}$ \cite{Brito:2013wya,Dias:2023ynv}. Upon imposing boundary conditions at the BH horizon and at spatial infinity, this system admits quasi-bound states with complex frequencies $\omega = \omega_R + i\,\omega_I$. When $\omega_I > 0$, the perturbation grows exponentially with time signaling an instability of the Kerr background---this occurs precisely for $\omega_R < m\,\Omega_H$ as in \eqref{eq:superradiance-condition}. 

The growth rate of the unstable perturbation is provided by $\omega_I$ and related to an instability timescale 
\begin{equation}
    \tau_\text{inst} = \frac{1}{\omega_I} \,,
    \label{eq:instability-timescale}
\end{equation}
with faster growth rates (larger $\omega_I$) leading to instabilities that take less time to set in. Given BH parameters $(M,\chi\equiv J/M^2)$, mode azimuthal number $m$ and boson mass $m_b$, one must solve the nonlinear eigenvalue problem to find the eigenfrequencies $\omega$ that obey the boundary conditions, and the corresponding instability timescale $\tau_\text{inst}$, 
\begin{equation}
    (M,\chi,m,m_b) \to \omega_I \to \tau_\text{inst} \,.
\end{equation}
This is the problem that was solved numerically in \cite{Dias:2023ynv} for the spin-2 case, for a generic BH spin $\chi$ and a large range of gravitational couplings $\alpha$, confirming prior results for the hydrogen-like modes and analysing in detail the special dipole ($m=1$) with the shortest instability timescale, also studied in \cite{Brito:2013wya}. For highly spinning BHs, the dipole mode has an instability timescale almost two orders of magnitude shorter than for any other superradiant mode.
This timescale can lead to a much wider forbidden region, spanning several orders of magnitude in BH masses---from stellar-mass to supermassive---for the same boson field mass $m_b$, and even become comparable to the typical BH ringdown \cite{Dias:2023ynv,Dias:2015wqa,Dias:2021yju,Dias:2022oqm,Berti:2009kk}. This would directly affect the post-merger phase through a dynamical reduction of the spin of the remnant during the ringdown, which could be seen in a gravitational waveform emitted by compact binary coalescences differing from the standard GR prediction.

These numerical results are accurately described by the polynomial fit \cite{Dias:2023ynv}
\begin{subequations}
    \begin{align}
        \label{eq:fit-omegaR}
        \frac{\omega_R}{m_b} &\approx \bigg(\sum_{i=0}^3 a_i\chi^i\bigg)\bigg[1+\alpha \sum_{i=0}^3 b_i\chi^i+\alpha^2 \sum_{i=0}^2 c_i\chi^i\bigg] \\
        \omega_I &\approx -\alpha^3\left(\omega_R - \Omega_H\right)\sum_{i=0}^2 d_i\chi^i \,, 
        \label{eq:fit-omegaI}
    \end{align}
    \label{eq:fits}
\end{subequations}

\noindent in the range $\alpha\in[0.05,0.8]$ and $\chi\in[0,0.99]$. The values of $a_i,b_i,c_i,d_i$ are given in \eqref{eq:fit-parameters}. We will use this numerical fit for the analysis in this work.

In the next subsection we explain how $\omega_I(M,\chi,m_b)$ can be compared with observational BH data to rule out bosons of a given mass $m_b$. 

\subsection{Constraints on bosonic field masses}

Due to superradiant instabilities, highly-spinning BHs satisfying condition \eqref{eq:superradiance-condition}  should lose energy and angular momentum in the presence of light bosonic fields, over a timescale $\tau_\text{inst}$ given by \eqref{eq:instability-timescale}. Whenever these timescales are much shorter than typical astrophysical timescales---including the time of accretion that could counteract this effect by spinning up the BH---the light boson leads to a forbidden region in the Regge plane (BH spin vs mass) for astrophysical BHs \cite{Arvanitaki:2010sy,Brito:2015oca,Baryakhtar:2017ngi,Brito:2017wnc,Brito:2017zvb,Ng:2019jsx,Fernandez:2019qbj,Stott:2020gjj,Unal:2020jiy,Unal:2023yxt}. If such a light field exists, a BH that spins fast enough to trigger an instability will start losing its spin, in a form of negative feedback that prevents BHs of a given mass from spinning any faster---we should therefore expect to see regions in the Regge plane that are completely unpopulated, corresponding to rotating BHs whose spins are too large for their masses. Any confident measurement of a non-zero spin of a BH that falls within these regions is enough to exclude a bosonic field with a mass that would trigger the instability.

Due to their smaller instability timescales, ultralight spin-2 fields of a given mass $m_b$ can be excluded by measurements of BHs with masses falling in a much wider range, making them easier to rule out \cite{Brito:2020lup,Dias:2023ynv}. Observational data for BHs spanning several orders of magnitude in mass, from stellar-mass to supermassive BHs, $M \in (1,10^{10})\,\Msol$, can exclude a wide range of spin-2 masses $m_b \in (10^{-23},10^{-11})$.

The most common astrophysical reference timescale is the Salpeter time \cite{Salpeter:1964kb,frank2002accretion},
\begin{equation}
    \tau_S \sim 4.5\times 10^{7}~\text{yr} \,,
\end{equation}
as a proxy for the minimum timescale for the BH spin to grow via gas accretion \cite{Brito:2015oca}. We will use this scale to determine whether the growth rate of an unstable mode is large enough to exclude a point $(M,\chi)$ on the Regge plane, by excluding the region 
\begin{equation}
    \tau_\text{inst}(M,\chi,m,m_b) < \tau_S \,.
    \label{eq:exclusion-condition}
\end{equation}

Before moving on to the main results of the paper, let us briefly comment on other signatures of superradiant instabilities besides the forbidden regions in the Regge plane. First of all, the boson condensate cloud formed around the BH is a continuous gravitational wave (GW) source, emitting a monochromatic signal with frequency
\begin{equation}
    f_{\rm GW} \sim 5 \,\text{kHz}\, \Big(\frac{m_b}{10^{-11}\text{eV}}\Big) \,,
\end{equation}
with the spin-2 cloud, in particular, emitting hexadecapolar radiation, rather than quadrupolar \cite{Brito:2015oca}. The boson condensate can also emit GWs in a beating pattern due to level transitions in its hydrogenic-like spectrum. Since many of these sources are expected to be too faint to detect individually, one could detect instead a stochastic background created through their incoherent superposition. Furthermore, nonlinear effects that become important when
the cloud reaches a critical mass can lead to a \emph{bosenova} explosion, whose signature would be a periodic emission of bursts \cite{Yoshino:2012kn}. 
Other signatures, of electromagnetic nature, include effects on BH shadows, or ``quasi-periodic-oscillations'' in the X-ray spectrum of accreting BHs, so-called BH lasers of frequency $f_{\rm laser}\sim\text{kHz}\,(m_b/10^{-11}\text{eV})$, to name a few. Finally, the boson cloud can also affect binary systems and pulsar timing, providing further avenues of detection (see \cite{Brito:2015oca} for an overview of all these signatures). Thus superradiant instabilities have a plethora of distinctive signatures that make rotating BHs extraordinary detectors of ultra-light bosons.

\subsection{Constraints on warped graviton KK towers}

We are now in a position to look for constraints on the masses of warped spin-2 KK towers given in \eqref{eq:RS-mass} due to superradiant instabilities of Kerr black holes, by imposing \eqref{eq:exclusion-condition} and using the fit for $\omega_I$ in \eqref{eq:fit-omegaI} from \cite{Dias:2023ynv}. These constraints can then be expressed in terms of the fundamental parameters $(k,r_c)$ of the warped compactification, as well as related to other physical properties such as, for instance, the maximum hierarchy between 5d and 4d scales (cf. \eqref{eq:physical-mass}).  

\begin{figure}[t]
\centering
\includegraphics[width=0.48\textwidth]{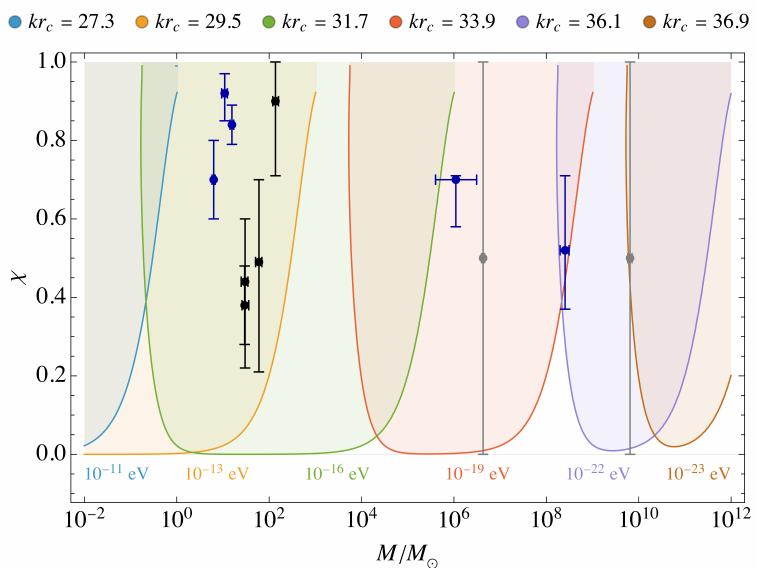}
\caption{BH spin-mass diagram with regions excluded by the superradiant instability of Kerr BHs against massive spin-2 fields (dominant $m=1$ mode). Figure reproduced from the results of \cite{Dias:2023ynv} through the polynomial fits \eqref{eq:fits} (for $\alpha<0.8$), with spin-2 masses translated to corresponding values of $kr_c$, with $k = \Mp$ for illustration.
}
\label{fig:Regge}
\end{figure}

We could consider, as in the case of fifth forces, taking the lightest KK mode in the tower, setting $m_b = m_1$ of \eqref{eq:RS-mass}, and studying the constraints for a single spin-2 field as in \cite{Dias:2023ynv}. In \figref{fig:Regge} we show the excluded regions in the $(M,\chi)$ plane for different values of $kr_c$ and fixing $k=\Mp$. Our choice of parameters $kr_c\in(27,37)$ effectively reproduces the results of \cite{Dias:2023ynv} for the mass range $m_b\in(10^{-23},10^{-11})$ eV. The plot also shows a small set of BH measurements for illustration, with black points corresponding to spin measurements from GW detections of compact binary coalescences by LIGO/Virgo/KAGRA \cite{LIGOScientific:2020ibl,LIGOScientific:2025slb}; blue points corresponding to electromagnetic estimates of the mass and spin of accreting stellar and supermassive BHs \cite{Liu:2008tk,Gou:2009ks,Brenneman:2011wz,Patrick:2012ua,Mastroserio:2020pat}; and gray points showing the supermassive BHs imaged by the Event Horizon Telescope, SgrA$^*$ and M87$^*$ (although an accurate spin measurement is still not available, these are expected to have moderate to large spins) \cite{EventHorizonTelescope:2019ggy,EventHorizonTelescope:2019dse,EventHorizonTelescope:2019pgp,EventHorizonTelescope:2022wkp,Tamburini:2019vrf,Davoudiasl:2019nlo}. 

\begin{figure*}[!t]
    \includegraphics[width=0.4\textwidth]{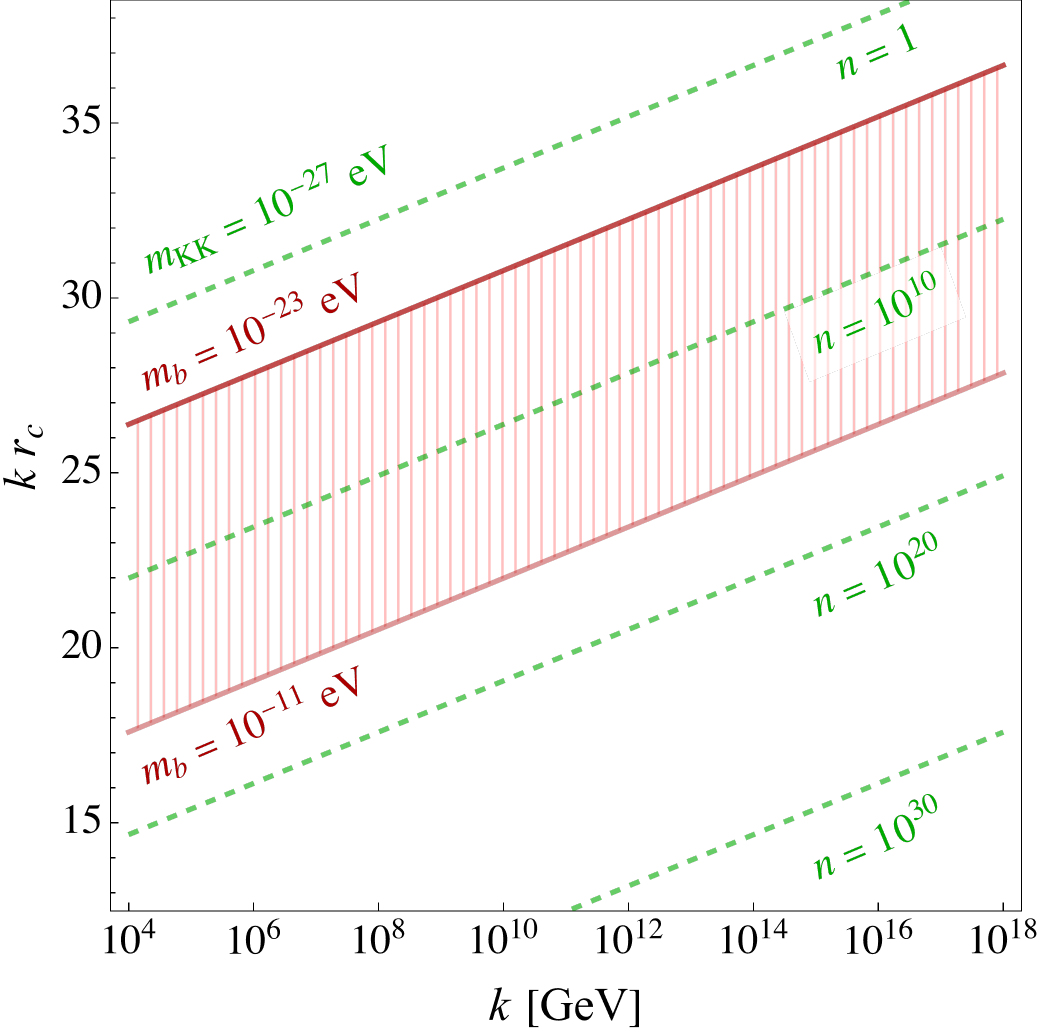}
    \hspace{3em}
    \includegraphics[width=0.4\textwidth]{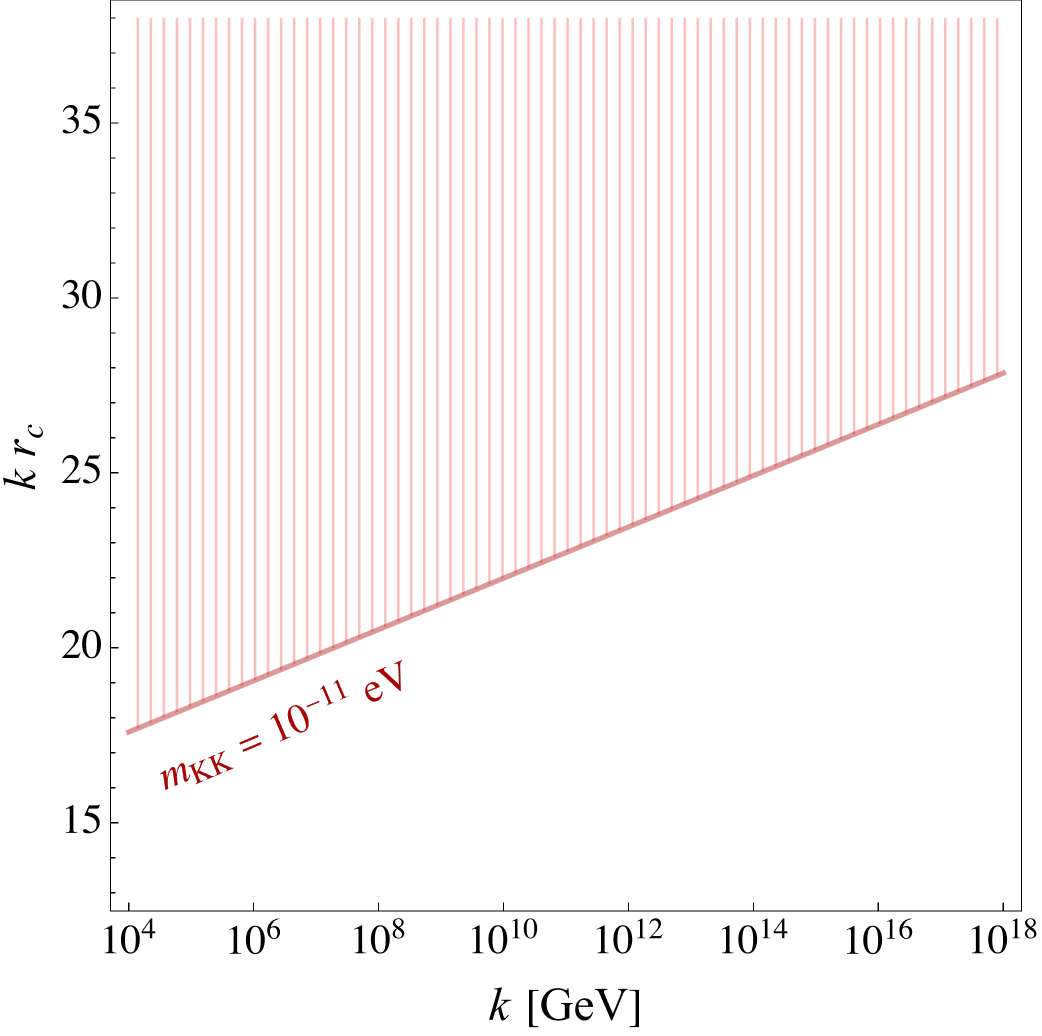}
    \caption{Constraints on the RS parameter space assuming BH spin measurements in the full range $M_{\rm BH}\in(1,10^{10})\Msol$. Each excluded mass $m_b$ rules out a line in the $(k,kr_c)$ plane; the shaded region between the solid lines corresponds to spin-2 masses in the range $m_b\in(10^{-23},10^{-11})$ eV. In the case of a Kaluza-Klein spin-2 tower, even if the lightest mode falls outside this region for being too light, some of the higher modes will be within this range.}
    \label{fig:RS-parameter-exclusion}
\end{figure*}
We see that confident spin measurements of BHs with masses $M\in(1,10^{10})\,\Msol$ have the potential to rule out spin-2 fields with masses in the range $m_b\in(10^{-23},10^{-11})$, which translates into the range 
\begin{equation}
    27 \lesssim kr_c \lesssim 37 \,,
\end{equation}
for $k=\Mp$. Since the superradiant constraints only depend on the mass of the field, there is a degeneracy between the parameters $k$ and $r_c$, with a given mass $m_b$ effectively corresponding to a line in $(k,kr_c)$ parameter space, which is ruled out for
\begin{equation}
    -23 \lesssim \log_{10}\frac{k}{\rm eV} + \log_{10}\gamma_1 - \frac{\pi\,kr_c}{\log 10} \lesssim -11 \,.
    \label{eq:line-krc-bounds}
\end{equation}
On the left panel of \figref{fig:RS-parameter-exclusion} we plot the region of parameter space that would be ruled out by BH spin measurements in the range $M\in(1,10^{10})\Msol$.

If there was only one spin-2 field of mass $m_b=m_1$, these would be the strongest constraints one could put on the parameters of the theory. 
However, a KK tower contains infinitely many spin-2 fields with masses arranged in a regular ladder that starts at $m_1 = m_{\rm KK}$. For the warped scenario we are considering, these masses are explicitly given in \eqref{eq:RS-mass}.
This means that the case of a KK tower is different in two ways. 

First, it may happen that the mass of the first KK mode is too small to trigger an instability of a BH with parameters $(M,\chi)$, but since the tower is infinite there will be a number of KK modes that fall within the superradiant window (cf. \figref{fig:RS-parameter-exclusion}).
Therefore, any KK tower with lightest mode mass $m_1\lesssim10^{-11}$ eV will contain spin-2 fields capable of triggering superradiant instabilities in the relevant range. As a result, we can exclude the much larger region
\begin{equation}
    \log_{10}\frac{k}{\rm eV} + \log_{10}\gamma_1 - \frac{\pi\,kr_c}{\log 10} \lesssim -11 
    \label{eq:RS-bounds-from-SR}
\end{equation}
on $(k,kr_c)$ parameter space. This stronger constraint is represented on the right panel of \figref{fig:RS-parameter-exclusion}. If we further require $k<\Mp$, so that the AdS$_5$ curvature scale is sub-Planckian and the 5d background solution is within the regime of validity of the 5d EFT \eqref{eq:4d-Planck-scale}, we find a direct bound on the warping strength
\begin{equation}
    kr_c \lesssim 28.5 \,,
\end{equation}
and consequently on the size of the extra dimension
\begin{equation}
    r_c \lesssim 28.5\Big(\frac{k}{\Mp}\Big)^{-1}\,\ell_P \,,
\end{equation}
in 4d Planck units. 

\begin{figure*}[t]
    \includegraphics[width=0.475\textwidth]{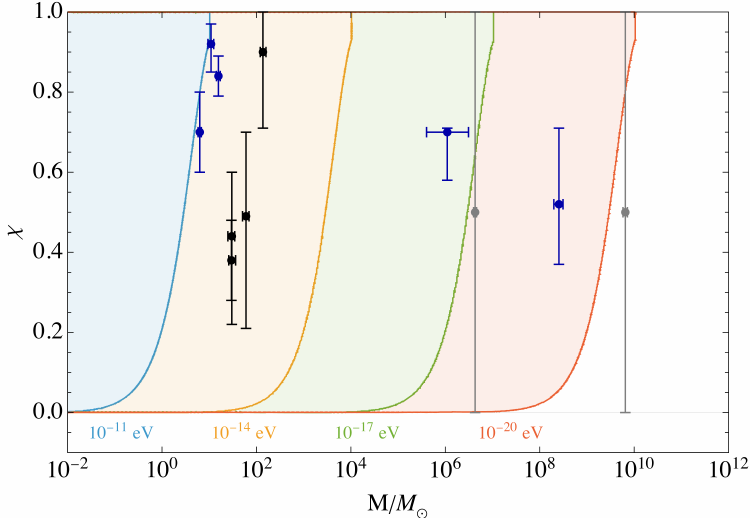}
    \hspace{1em}
    \includegraphics[width=0.475\textwidth]{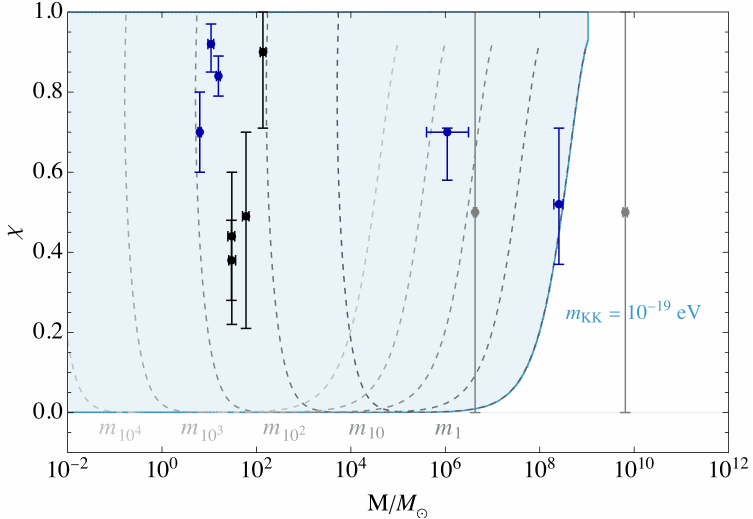}
    \caption{BH spin-mass diagram with regions excluded by the superradiant instability of Kerr BHs against a massive spin-2 KK tower (dominant $m=1$ mode). \emph{(Left)} Exclusion regions for different values of the KK mass gap, showing how the infinite tower excludes all BH masses lower than some value (cf. with single massive spin-2 state, where only a mass range  was excluded). \emph{(Right)} Exclusion region for $m_{\rm KK}=10^{-19}$ eV overlaid with the contours of regions excluded individually by a few different modes in the tower; the collective contribution of the full tower is equivalent to the union of the individual contributions of each mode.}
    \label{fig:KK-Regge}
\end{figure*}

The second effect of considering a KK tower is due to the multiplicity of fields. When a KK mode in the tower is light enough to trigger a superradiant instability, at least the first few modes that follow are also likely to fall within the instability mass range. In this scenario, there is a potentially large number of spin-2 fields that can make the Kerr background unstable individually. Thus we should expect the most stringent constraints on the fundamental parameters to follow from taking into account the collective effect of the tower. 

This is similar to the study of superradiant instabilities in the context of the axiverse, in which there is a large number of ultra-light axions \cite{Arvanitaki:2009fg,Arvanitaki:2010sy,Stott:2017hvl,Stott:2018opm,Stott:2020gjj,Caputo:2025oap,Leedom:2025mlr}. As for the axiverse case \cite{Stott:2018opm,Stott:2020gjj}, we will assume that there are no interactions between the different spin-2 states and that the backreaction on the Kerr background due to the bosonic cloud is negligible. In these conditions, the total instability growth rate $\omega_I^{\rm tot}$ for a BH with parameters $(M,\chi)$ due to the spin-2 tower is given by the sum of individual rates
\begin{equation}
    \omega_I^{\rm tot} = \sum_{n} \omega_I^{(n)} \,,
    \label{eq:total-growth-rate}
\end{equation}
over all levels $n$ whose masses $m_n$ are in the right range to trigger an instability for this particular BH, i.e. with $\omega_R^{(n)} < \Omega_H$. 

On the left panel of \figref{fig:KK-Regge} we show the excluded regions in the BH spin-mass diagram, where $\omega_I^{\rm tot} > \tau_S^{-1}$, for different values of $m_{\rm KK}$. As expected, taking into account the KK tower extends the excluded regions all the way to smaller values of $M$. This is consistent with our conclusion above that KK towers with mass gap $m_{\rm KK} \lesssim 10^{-11}$ eV can be ruled out even with BHs of stellar mass. 

On the other hand, overlaying the contours for the regions that are excluded individually by a few different modes in the tower suggests that,
at least within the regime of validity of our analysis, the collective contribution of multiple KK modes does not affect significantly the instability rate of a given BH with parameters $(M,\chi)$. In other words, the individual contribution of \emph{some} mode in the tower is enough to exclude that point in the spin-mass diagram.

\section{Discussion}
\label{sec:conclusions}

Superradiant instabilities of rotating BHs provide a powerful way of detecting or ruling out ultralight bosons, regardless of how weakly they couple to SM fields, making these BHs particularly useful to probe fields with extremely weak couplings that may easily evade fifth force constraints. On the other hand, given the relevant mass range of astrophysical BHs, $M\in(1,10^{10})\Msol$, this method only probes ultra-light fields with masses $m_b\lesssim 10^{-11}$ eV. 

From the point of view of string theory, the most obvious targets are perhaps the ubiquitous light moduli of a string compactification; among them can be numerous axions, whose masses may be extremely light due to their shift symmetries that protect them against mass generating effects \cite{Arvanitaki:2009fg,Arvanitaki:2010sy,Stott:2017hvl,Stott:2018opm,Stott:2020gjj,Caputo:2025oap,Leedom:2025mlr}. Another potential target arises when one considers warped compactifications, with exponentially light Kaluza-Klein towers of massive states. A prototypical toy model of a warped compactification is given by the 5-dimensional Randall-Sundrum model \cite{Randall:1999ee,Randall:1999vf} that we reviewed in Section~\ref{sec:RS}. 

Among the KK towers there is always one whose states are massive spin-2 fields in four dimensions. It has been recently shown \cite{Dias:2023ynv} that massive spin-2 fields lead to the strongest superradiant instabilities of Kerr BH backgrounds, with instability times much shorter than those of scalar and vector perturbations. As a consequence, ultralight spin-2 fields can be ruled out by spin measurements of BHs in a large portion of the Regge plane, and even affect the ringdown behaviour of a BH merger \cite{Dias:2023ynv}. Thus, rotating BHs can potentially rule out a large region of parameter space of warped compactifications.

We have studied the special dipolar $(m=1)$ unstable mode \cite{Brito:2013wya,Dias:2023ynv} in the context of a warped compactification of one extra dimension, using the Randall-Sundrum scenario as a concrete example. We have shown that a KK tower can trigger an instability for all BH masses $M<M_{\rm max}$ for any given spin $\chi$, in contrast with the BH mass window that becomes unstable in the presence of a single spin-2 field. In particular, spin measurements of BHs with $M\gtrsim \Msol$ put a bound on the warp factor parameters $k\,r_c\lesssim 28.5$, which becomes tighter when $k$ is parametrically smaller than $\Mp$; for example, if $k \sim 100$ TeV, the bound on the warping becomes $k\,r_c\lesssim 18.8$.  

Although one might expect the high multiplicity of KK modes to   exclude a larger region in the Regge plane due to a significant increase of the instability rate, we can see that this is not the case by comparing the region excluded by the tower with the exclusion regions of individual modes (right panel of \figref{fig:KK-Regge}). We see that the tower excluded the same region as the union of the individual modes, but not more. This is a consequence of the strength of the instability triggered by the dipolar mode of a single spin-2 field, which excludes spins $\chi$ essentially all the way to zero for some BH and spin-2 masses. One can then only hope to see this effect on the right-end of the exclusion region, since this is where we find BHs that are safe from the instabilities triggered by any individual mode. However, on this end we also have very low multiplicity, so that the collective contribution of modes falling in the superradiant window is still not enough to lower the instability time significantly below the Salpeter time. Although the hydrogenic modes $(m=2)$ are subdominant with respect to the dipolar one, we expect this effect to be more relevant for those, since the corresponding excluded regions are smaller.

To a first approximation, these bounds can be mapped to well motivated warped string backgrounds, such as the Klebanov-Strassler throat \cite{Klebanov:2000hb}. In this context, we get bounds on a combination of parameters that includes the volume of the compact space, the string coupling and the flux numbers sourcing the warping \eqref{eq:KS-mass}. Comparing \eqref{eq:KS-mass} and \eqref{eq:RS-mass}, we find the relations 
\begin{align}
    \frac{k}{\Mp} = \frac{1}{M\mathcal{V}^{1/3}} \,,
    \quad kr_c = \frac{4K}{3g_sM} \,, 
\end{align}
which we can use to translate the bounds in \eqref{eq:RS-bounds-from-SR} into bounds on these parameters, 
\begin{equation}
    \log_{10}\frac{1}{M\mathcal{V}^{1/3}} + \log_{10}\gamma_1 - \frac{4\pi\,K}{3g_sM\,\log 10} \lesssim -39 \,,
    \label{eq:KS-bounds-from-SR}
\end{equation}
with $\gamma_1 \approx 2.58$ for the warped deformed conifold \cite{ValeixoBento:2022qca}. Note that the volume $\mathcal{V}$ is expressed in string units, i.e. $\mathcal{V} = V/\ell_s$, with $\ell_s = m_s^{-1}$ and $m_s = \tfrac{g_s}{\sqrt{4\pi\mathcal{V}}}\Mp$. Control of the supergravity approximation requires $\mathcal{V}> 1$, which leads to the bound
\begin{equation}
    \frac{4K}{3g_sM} \lesssim 28.5 \,,
\end{equation}
and constrains the strength of the warping to 
\begin{equation}
    e^{2A_{\rm tip}} \gtrsim \frac{2.4\times 10^{-5}}{(g_s M)\mathcal{V}^{1/3}} \,. 
    \label{eq:}
\end{equation}
In de Sitter vacua that involve the uplfit of an AdS vacuum by an $\rm\overline{D3}$--brane, such as in KKLT or LVS, the brane sits at the tip of the conifold with a contribution to the potential suppressed as
\begin{equation}
    V_{\rm\overline{D3}} \sim \frac{e^{4A_{tip}}}{\mathcal{V}^{4/3}} \gtrsim \frac{10^{-14}}{K\,\mathcal{V}^{4/3}} \,,
\end{equation}
after using the bound from superradiance in the last inequality. This puts an upper bound on the warping suppression of the uplift, which may not be enough to avoid the volume runaway depending on the magnitude of the remaining terms in the potential (i.e. on the depth of the pre-uplift AdS vacuum). 

Finally, the same way the instabilities of higher-dimensional black strings and branes of the form Kerr$_4\times\mathbb{R}^p$ could be interpreted as superradiant instabilities in 4d triggered by KK modes, one could interpret the superradiant instabilities due to the warped KK tower as an instability of black strings and branes in warped higher-dimensional backgrounds, such as the AdS$_5$ of RS. It would be interesting to explore this higher-dimensional point of view in more detail. 

\appendix
\section{Superradiant instabilities in Kerr BHs}
\label{ap:Kerr}

In the usual Boyer--Lindquist coordinates, the Kerr background is described by the metric
\begin{align}
    \label{eq:kerr-line-element}
    ds^2 =& -\frac{\Delta(r)}{\Sigma(r,\vartheta)}\big[dt - a\sin^2\vartheta d\varphi\big]^2
    +\frac{\Sigma(r,\vartheta)}{\Delta(r)} dr^2  \\
    &+ \Sigma(r,\vartheta) d\vartheta^2
    + \frac{\sin^2\vartheta}{\Sigma(r,\vartheta)}\big[a\,dt - (r^2 + a^2) d\varphi\big]^2 \,, \nonumber
\end{align}
where $(r, \vartheta, \varphi)$ are oblate spheroidal coordinates, $a = J/M$, and
\begin{subequations}
    \begin{align}
    \Sigma(r,\vartheta)&\equiv r^2 + a^2\cos^2{\vartheta} \,, \\
    \Delta(r)&\equiv r^2 - 2M r + a^2 \,.
\end{align}
\end{subequations}
The event horizon corresponds to a null hyper-surface with $r=r_+$, where $r_+$ is the largest positive real root of $\Delta(r)$. In the (non-extremal) Kerr background there is a distinct second root, $r_i < r_+$, corresponding to a Cauchy horizon, and crucially an ergoregion bounded by the ergosphere at 
\begin{equation}
    r_\text{ergo} = M\,\big[1 + \sqrt{1-\chi^2\cos^2\vartheta}\big]\,,
\end{equation}
where $\chi\equiv a/M \leq 1$ is the dimensionless BH angular momentum. It is precisely the ergoregion that allows for energy extraction from the vacuum and the process of superradiance \cite{Brito:2015oca}.

Here we are interested in massive spin-2 fields propagating on this Kerr BH background; in other words, on massive spin-2 perturbations $H_{\mu\nu}$ of the metric \eqref{eq:kerr-line-element} that are described by the equations
\begin{equation}
    \Box H_{\mu\nu} + 2R_{\mu\nu\rho\sigma}H^{\rho\sigma} - \mu^2 H_{\mu\nu} = 0 \,,
    \label{eq:spin-2-field-equations}
\end{equation}
subject to the constraints
\begin{equation}
    \nabla^\mu H_{\mu\nu} = 0\,,
    \quad H^\mu_{\phantom{\mu}\mu} = 0 \,,
    \label{eq:spin-2-field-constraints}
\end{equation}
where $\Box\equiv\nabla^\mu\nabla_\mu$ is the d'Alembert operator in 4d and $R_{\mu\nu\rho\sigma}$ the Riemann tensor of the background. The field $H_{\mu\nu}$ propagates five physical degrees of freedom; note that the constraints \eqref{eq:spin-2-field-constraints} are \emph{not} gauge choices \cite{Brito:2013wya}. The equations \eqref{eq:spin-2-field-equations} are not separable on a Kerr background \cite{Brito:2020lup,Dias:2023ynv} and have thus been solved approximately in the $\alpha = G\mu M\ll 1$ limit using matched asymptotics \cite{Brito:2020lup} or fully numerically \cite{Dias:2023ynv}.

Crucially, the mass term of the field $H_{\mu\nu}$ can confine perturbations that are amplified through superradiance, leading to an exponential growth. This confining mechanism can be seen in the form of quasi-bound states of an effective potential generated by the field mass \cite{Brito:2015oca}. 

Since $\partial_t$ and $\partial_{\varphi}$ are Killing vector fields of \eqref{eq:kerr-line-element}, the spin-2 field perturbations admit a natural decomposition along those directions,
\begin{equation}
    H_{\mu\nu} (t,r,\vartheta,\varphi) = e^{-i\omega t}e^{im\varphi}\tilde{H}_{\mu\nu}(r,\vartheta) \,,
\end{equation}
with $m$ the azimuthal number and $\omega$ the frequency of the perturbation. The resulting equations for $\tilde{H}_{\mu\nu}$ are a coupled system of PDEs that constitute an eigenvalue problem for the complex frequency $\omega = \omega_R + i\,\omega_I$, for each mode $m$, once suitable boundary conditions are imposed. 

More precisely, we must provide boundary conditions at the horizon, at spatial infinity and at the poles of the $S^2$. At spatial infinity and at the poles, we impose regularity of the perturbation \cite{Berti:2005gp}, which eliminates any solution that grows unbounded in these limits. Considering the causal structure of the Kerr BH, we must have \emph{purely ingoing} boundary conditions at the horizon, i.e. no waves can come out of the BH.
This requirement is what renders the boundary-value problem non-Hermitian, making the eigenfrequencies complex, and leading to the growth or decay of the perturbations,
\begin{equation}
    H_{\mu\nu}\,\, \propto\,\, e^{\omega_I t}\cos(\omega_R\,t)\,\tilde{H}_{\mu\nu} \,.
\end{equation}
Therefore, when $\omega_I>0$ the perturbation grows exponentially in time, signalling an instability of the background, with an instability timescale given by $\tau = 1/\omega_I$. Note also that $\omega_R$ determines the oscillating frequency of the perturbation and will determine which modes have an angular phase velocity that is lower than the angular velocity of the background, 
\begin{align}
    \Omega_H = \frac{1}{2M}\,\frac{\chi}{1+\sqrt{1-\chi^2}} \,.
\end{align}
This determines the superradiance condition
\begin{equation}
    \omega_R < m\,\Omega_H \,. 
\end{equation}
When this condition is saturated, $\omega_I = 0$ and true bound states are possible, which can lead to hairy BH solutions with scalar clouds \cite{Herdeiro:2014goa}.

In \cite{Dias:2023ynv}, the eigenvalue problem was solved using a Newton-Raphson root-finding algorithm \cite{Dias:2015nua}. The shortest instability timescale is found to be associated with the special dipolar mode $(m=1)$ already studied in \cite{Brito:2013wya}. These numerical results are accurately described by the polynomial fit \cite{Dias:2023ynv}
\begin{subequations}
    \begin{align}
        \frac{\omega_R}{\mu} &\approx \bigg(\sum_{i=0}^3 a_i\chi^i\bigg)\bigg[1+\alpha \sum_{i=0}^3 b_i\chi^i+\alpha^2 \sum_{i=0}^2 c_i\chi^i\bigg] \\
        \omega_I &\approx -\alpha^3\left(\omega_R - \Omega_H\right)\sum_{i=0}^2 d_i\chi^i \,, 
    \end{align}
\end{subequations}
with 
\begin{align}
    a_i &= (0.73, -0.05, 0.15, -0.12) \,,\nonumber  \\ 
    \label{eq:fit-parameters}
    b_i &= (-1.21, 0.68, -0.55, 0.61) \,, \\
    c_i &= (0.69, -0.58, -0.11) \,, \nonumber \\
    d_i &= (1.47, 1.86, -2.75) \,. \nonumber 
\end{align}
In the unstable regime, these fits are accurate within 2\% and 80\%, respectively, in the range $\alpha\in [0.05, 0.8]$ and $\chi\in [0, \approx 0.99]$ \cite{Dias:2023ynv}. 

\acknowledgments
We are indebted to
Oscar Dias, 
Pedro Fernandes,
Edward Hardy,
Miguel Montero,
Susha Parameswaran 
and Salvatore Raucci 
for valuable discussions.
We acknowledge the support of an Atraccion del Talento Fellowship 2022-T1/TIC-23956 from Comunidad de Madrid, in the early stages of this project, as well as a fellowship from Fundación Ramón Areces, and the Spanish State Research Agency (Agencia Estatal de Investigacion) through the grants IFT Centro de Excelencia Severo Ochoa CEX2020-001007-S, PID2021-123017NB-I00, and Europa Excelencia EUR2024-153547.

\bibliography{paper.bib}
\end{document}